\documentclass[twocolumn,aps,amsmath,amssymb,showpacs]{revtex4}

\usepackage{graphicx}
\usepackage{epsfig}

\begin{document}

\title{Statistics of football dynamics}
\author{R. S. Mendes}
\affiliation{Departamento de F\'{\i}sica,
Universidade Estadual de Maring\'a, Av. Colombo 5790, 87020-900, Maring\'a-Paran\'a, Brazil}
\author{L. C. Malacarne}
\affiliation{Departamento de F\'{\i}sica,
Universidade Estadual de Maring\'a, Av. Colombo 5790, 87020-900, Maring\'a-Paran\'a, Brazil}
\author{C. Anteneodo}
\affiliation{Departamento de F\'{\i}sica, Pontif\'{\i}cia Universidade
Cat\'olica do Rio de Janeiro,  CP 38071, 22452-970, Rio de Janeiro, Brazil}


\begin{abstract}

We investigate the dynamics of football matches.
Our goal is to characterize statistically the temporal sequence of ball
movements in this collective sport game, searching for traits of complex behavior.
Data were collected over a variety of matches in South American,
European and World championships throughout 2005 and 2006.
We show that the statistics of ball touches presents power-law tails and
can be described by $q$-gamma distributions.
To explain such behavior we propose a model that provides information
on the characteristics of football dynamics.
Furthermore, we discuss  the statistics of  duration of out-of-play intervals,
not directly related to the previous scenario.

\end{abstract}


\maketitle

\section{Introduction}

The statistical analysis of physical systems
has been fundamental for the identification of underlying mechanisms.
This kind of analysis has also
found applications in other fields, such as biology, economics and
social sciences, with a reciprocal feedback for understanding
physical systems.
In any context, when dealing with complex systems, where
individual, collective and aleatory features may be present, a central interest
is to trace general statistical properties of the dynamics.
This suggests a direction to investigate, amongst other human activities,
collective sports such as the most popular one: football.
In fact, in football matches, player actions range from
elementary individual reactions to elaborated strategies
involving several players,  motivating the
search for traits of complex behavior.

Amongst the diverse football games, we restrict our study to
male official association football (soccer).
Because of its popularity and widespread diffusion in the media,
an abundant source of observational data is accessible.
Previous works have dealt mainly with macroscopic features measured over
{\em ensembles} of matches
(cups or championships) \cite{goles,newman,leagues,onody}, such as the statistics of
goals.
Meanwhile, the present goal is to characterize
a microscopic  dynamics throughout each match.
From a related perspective, the detection of
temporal patterns of behavior has been pursued before \cite{patterns}.
Differently, in this work, we analyze
the stream of ball events throughout a match from a statistical point of view.
We focus on the
temporal aspects, without taking into
account any spatial counterpart, e.g., players or ball trajectories in space.

The results we report in this paper derive from information collected
over twenty six matches in South American and
European championships in 2005-2006 (Table \ref{tableone}).
Temporal series were obtained from the sequence
of touches in each football match.

After the presentation of collected data in Sec. II, we  expose in Sec. III our modeling
of the statistics of times between touches, starting with simple exponential
distributions, refining the description through gamma distributions and finally
through generalized gamma distributions.
Basically, we show that a non-stationary Poisson process allows
to describe with success the main statistical properties.
Before making final observations (Sec. V), we also discuss the statistics
of in and out-of-play intervals (Sec. IV).
On one hand our results show the possibility of
applying statistical physics methods to study collective
aspects of sports such as football, evidencing interesting features.
On the other, we exhibit a concrete example where the
mechanisms of the ``superstatistics'' recently proposed
by Beck and Cohen\cite{beck1}, in connection with
Tsallis statistics\cite{tsallis}, and
where a generalized gamma distribution ($q$-gamma) plays
a central role, apply.
Therefore, our results may provide insights on a more general context.

\section{Data acquisition and preliminary analysis}

Data were acquired from TV broadcasted matches,
with the aid a computer program that records (with precision of $10^{-2}$s)
the instants of time  at which predetermined keys are clicked.
Thus, time was recorded by clicking when a given player action takes place.
Considering that the average human reaction time is about 0.1-0.2 s, this may lead to a
systematic error of that order in data recording. However,  in the present study we are
interested in time differences, therefore, any systematic error will be reduced. Moreover,
some of the matches were first recorded and then played in slow motion for data collection,
showing no significant differences from those obtained in real time.
Also, in order to check that other biases were not being introduced by the observer, the authors
independently obtained the data for some of the matches and verified that the
corresponding distributions essentially remained the same.

For each match considered, we monitored the
occurrence of touches (kicks, headers, shots, throw-ins,  etc.)  that change the
player in possession of the ball.
That is, touches from one player to himself were not taken into account.
The instants of occurrence of ball touches were recorded without
making distinction on the type of touch, nor on the subsequent movement of
the ball (rolls, flies, etc.), nor on whether it was intentional or
accidental. Moreover, touches were not distinguished by teams.
We also recorded
the time at which each sequence of touches ends, that typically
corresponds to the instant when the whistle is blown.

\begin{figure}[ht]
\begin{center}
\includegraphics*[bb=120 530 520 750, width=0.45\textwidth]{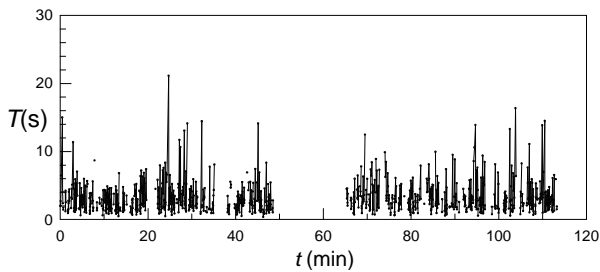}
\caption{ Time series of $T$ (time that elapses between consecutive touches).
The match is Chelsea $\times$ Barcelona in 22/Feb./2006 (UEFA Champions League).
Symbols are joined by linear segments whenever there is no interruption between
touches.
}
\label{fig:series}
\end{center}
\end{figure}

First, let us consider the variable $T$ corresponding to the time that elapses
between two consecutive touches occurring without interruption of the match
(inter-touch time).
A typical time series of inter-touch times in a match is exhibited in Fig.~\ref{fig:series}.
This plot manifests the discontinuous nature of football
activity where the sequences of touches are interrupted
by events such as the ball leaving the field, player fouls, defective ball,
external interference or any other reason to stop the game.  Then, time series
are characterized by sequences of ball-in-play fragments.

\begin{figure}[hb]
\begin{center}
\includegraphics*[bb=100 450 550 770, width=0.45\textwidth]{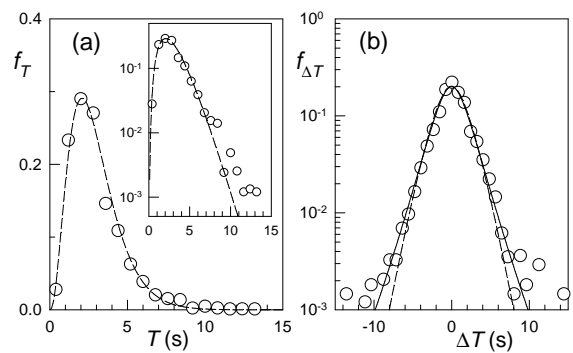}
\caption{ Normalized histograms of (a) inter-touch times and
(b) their increments, for the match of Fig.~\ref{fig:series}.
The inset in (a) is a semi-log representation of the same data plotted in the main frame.
Dashed lines represent: (a) Gamma PDF given by Eq. (\ref{gama}) with
parameters $\beta=3.43\pm0.17$ and $\tau/$s$=0.85\pm0.05$, and (b)
the corresponding PDF of increments given by Eq.~(\ref{n3}).
The full line in (b) corresponds to Eq.~(\ref{teo2}).
}
\label{fig:histos}
\end{center}
\end{figure}

Typical histograms of inter-touch times $T$ and increments $\Delta T$ (between consecutive
inter-touch times) are presented in Fig.~\ref{fig:histos}(a-b).
Unless otherwise stated to the contrary, in this and subsequent analyses,  touches of goal-keepers
were not considered, because of the singular role in the game.
These plots, built for the match of Fig.~\ref{fig:series}, are quite similar to
those observed for other matches.
In all cases the decay of the probability density function (PDF) of inter-touch times is
approximately exponential and the PDF of
increments presents a ``tent'' shape
in the semi-log representation, suggesting a double exponential
decay.

\begin{figure}[ht]
\begin{center}
\includegraphics*[bb=100 420 530 780, width=0.45\textwidth]{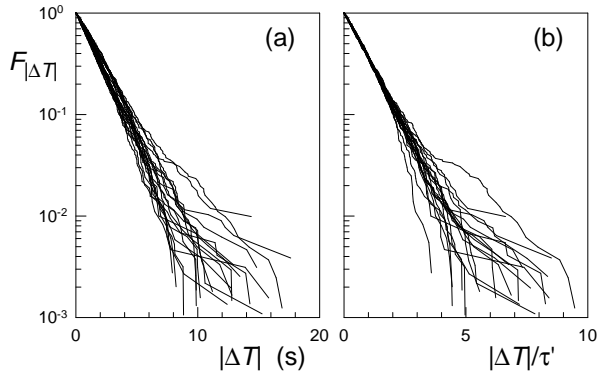}
\caption{Cumulative distribution of increments.
In (a) $F_{ |\Delta T|}(t)=Prob(|\Delta T|\ge t)$  was computed for each of the twenty
matches  listed in Fig.~\ref{fig:pars}. In (b), increments
are scaled by the respective exponential decay time $\tau'$
obtained from the exponential fit to each curve in (a), performed for values
up to approx. $2\tau'$ only.
}
\label{fig:cum}
\end{center}
\end{figure}

Histograms of the increments, built for several matches, are
exhibited in Fig.~\ref{fig:cum}(a).
Cumulative distributions were considered to reduce fluctuations.
In order to quantitatively characterize each match,
we calculated the exponential decay time $\tau'$,
since in all cases the initial decay is exponential.
Calculation of $\tau'$ was performed through an exponential fit
to the cumulative distributions of increments over the interval $\Delta T <4s$.
The fitting procedure, that will denominate WLS along the paper,
was weighted least squares, with weights
$w_i=1/|y_i|$ attributed to each data point $(x_i,y_i)$, and
was performed by means of commercial software Origin.
Fig.~\ref{fig:cum}(a) already exhibits qualitatively the narrow statistical diversity
found from one match to another, concerning the most frequent events.
Constraints, such as rules of the game and human effort limitations, confine
single realizations to a narrow spectrum.
In fact, values of $\tau'$ fall within the narrow interval $\tau'=(1.89\pm0.18$)~s.
Collapse of the re-scaled data is observed up
to $|\Delta T| \simeq 2\tau'$ only (Fig.~\ref{fig:cum}(b)).
Naturally,
deviations from the mean behavior occur for large $\Delta T$ because
events are rare.

\section{Modeling inter-touch time statistics}
Let us consider the random
variable $X_t$ representing the number of occurrences during the period
$[0,t)$. If the stochastic process $\{X_t, t>0\}$ were purely Poissonian
(as commonly considered for arrival time statistics),
with expected rate $1/\tau$,
then the PDF of inter-touch times should be
the exponential $f_{T}(t)={\rm e}^{-t/\tau}/\tau$ \cite{probs}.
Moreover,  given two independent variables $T_1$, $T_2$ with the
same exponential distribution,
the increment $\Delta T=T_2-T_1$ has the so called double exponential
or Laplace PDF

\begin{equation}
f_{\Delta T}(t) = \frac{1}{2\tau}{\rm e}^{-|t|/\tau} .
\end{equation}

Although the distribution of increments $\Delta T$ is in good accord with
a Laplace PDF,
at least for central values, the distribution
of inter-touch times $T$ is clearly not a pure exponential (see Fig.~\ref{fig:histos}(a)).
Therefore, the exponential, with one single fitting parameter,
constitutes a very coarse model for the histograms of inter-touch times.

Instead, the time interval between touches can be thought to be
composed,  in average, of a certain number $\beta$ of independent phases.
If each of the phases is
$\tau$-exponentially distributed, then one obtains the Erlang distribution:
\begin{equation}\label{gama}
f_{T}(t)=\frac{1}{\tau\Gamma(\beta)}(t/\tau)^{\beta-1}{\rm e}^{-t/\tau}
\equiv \Gamma_{\beta,\tau}(t),
\end{equation}
defined for $t\geq 0$.
This PDF is also known as gamma distribution, for real $\beta\ge 1$.
In the particular case $\beta=1$, one recovers the pure exponential distribution.
However,  in the vicinity of the origin the PDF of inter-touch
times  has a shape compatible with $\beta>1$.
In fact, very short inter-touch times are not frequent
since players are not typically so close to each other.
Very long inter-touch times are also scars since teams
dispute ball possession almost all the time.
Moreover, let us remark that the Erlang distribution is commonly used  for modeling
the distribution of times to perform some compound task,
such as repairing a machine or completing a customer service \cite{task}.
Also in the present case, when a player receives the ball,
it is common that he executes more than one task,
such as, keeping possession
of the ball, avoiding opponents, and passing the ball to another member of his team.
In what follows we do not restrict $\beta$ to take integer values. Then $\beta$
can be interpreted as an average number of phases.

Fig.~\ref{fig:histos}(a)  shows the results of WLS fitting
the gamma PDF  to the numerical histograms of inter-touch times.
We observe a clear improvement
in the description of the statistics of inter-touch times,
in comparison with the pure exponential model, for small and moderate times.
Assuming a gamma distribution, parameters $\beta$ and $\tau$  were determined
by means of a least square fit to empirical PDFs, using statistical weights
in order to ponder the tails of the distributions. Even so,
although a good description of numerical histograms is obtained for small
and moderate values, there are important deviations at the tails.
This also suggests that the gamma distribution may not be a
very good model for the present data.

\begin{figure}[hb!]
\begin{center}
\includegraphics*[bb=140 390 500 615, width=0.45\textwidth]{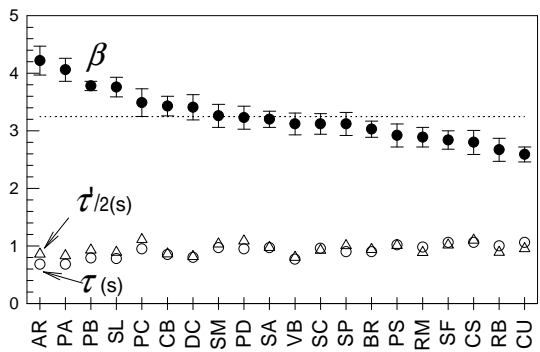}
\caption{ Parameters $\beta$ and $\tau$, obtained from the WLS fit of the
gamma distribution to the histograms of inter-touch times, for the matches indicated
in the abscissa axis (see Table~\ref{tableone} for letter code).
Parameter $\tau'$, obtained from cumulative
histograms of increments is also exhibited. Matches were ranked by decreasing $\beta$.
}
\label{fig:pars}
\end{center}
\end{figure}

Fig.~\ref{fig:pars} exhibits the values of parameters $\beta$ and $\tau$
for several matches, together with the value of $\tau'$ (for the
cumulative distribution of increments).
Values of $\tau$ are found within the interval $[0.68,1.06]$, with average
value $0.91$, while $\beta$ values fall within the interval $[2.6,4.2]$, with mean $3.24$.
There is a tendency that faster games (smaller $\tau$) are characterized
by a larger $\beta$, as clearly observed in Fig.~\ref{fig:pars}.
There is also a trend that more decisive matches
or matches played by highly ranked teams have smaller $\tau$ and larger $\beta$,
 e.g., {\sf PA, SL} are final matches,
 {\sf AR, CB} are usual matches of the UEFA champions league.
These tendencies are expected because in such matches players usually save no efforts
and strategies are more elaborated.
If we assume that successive inter-touch times are independent  identically
gamma-distributed variables,
then, the PDF of increments  becomes
\begin{equation}  \label{teo}
f_{\Delta T}(t)=\frac{{\rm e}^{-|t|/\tau}}{\tau^{2\beta} [\Gamma(\beta)]^2}
\int_0^\infty [x(|t|+x)]^{\beta-1}{\rm e}^{-2x/\tau} \,dx\;.
\end{equation}

Notice that this function
has the same asymptotic behavior as the Laplace PDF but it is
smoother at the origin.
In particular, for $\beta=3$ (integer value closer to the average one), one gets
\begin{equation} \label{n3}
f_{\Delta T}(t) =\frac{1}{16\tau} \left[
3+(3+|t|/\tau)|t|/\tau
\right]{\rm e}^{-|t|/\tau}.
\end{equation}

Once obtained the values of parameters $(\beta,\tau)$ for the
distribution of inter-touch times,
they were used to predict the PDF  of
increments by means of Eq. (\ref{teo}).
From the results, shown in Fig.~\ref{fig:histos}(b),
one concludes that the prediction of the PDF of increments, assuming independence of
consecutive inter-touch times, is satisfactory except at the tails.

Let us discuss some points that may be responsible for deviations from
the simple Poissonian framework.
First, we investigated the assumption of independence of occurrences in
non-overlapping intervals.
We calculated  auto-correlation
functions both for variables $T$ and $\Delta T$, taking into account
the intrinsically discontinuous nature of the time series. Hence, only
pairs of times belonging to the same sequence of touches were considered.
No significant correlations were detected in the series of $T$, although
the series of $\Delta T$ typically presents traits of antipersistence.
Hence, despite  the existence of strategies and patterns
involving several players (some in cooperation and others in opposition),
the lack of significant correlations indicates that memory effects in the
succession of touches are very weak.

\begin{figure}[ht]
\begin{center}
\includegraphics*[bb=70 350 540 720, width=0.45\textwidth]{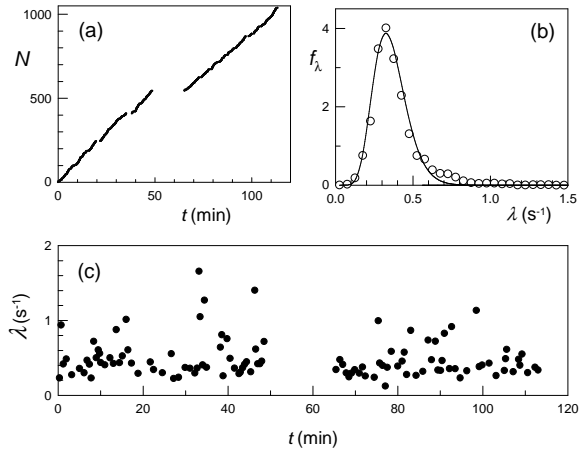}
\caption{
Number of events as a function of time (a), for the mathc of
Fig.~\ref{fig:series}.  Histogram of local rates $\lambda$ (b).
Solid lines represent WLS fits of a gamma distribution
$\Gamma_{\alpha,\kappa}$ with $(\alpha,\kappa/$s$^{-1})=$ (11.1,0.032)
Local rate $\lambda$ as a function of time (c). }
\label{fig:rates}
\end{center}
\end{figure}

Another possibility for the failure of the simple Poissonian picture
concerns stationarity  and homogeneity.
In order to investigate this aspect, we analyzed for each match
the number of events $N$ as a function of time, as
illustrated in Fig.~\ref{fig:rates}.
Beyond the discontinuity of the time series, temporal inhomogeneities
throughout a match are common.
First of all, average rates computed over each half of the matches
are almost always different. In most of the cases, these average rates
are larger in the first half, as soon as
players are usually more tired in the second half of a match.
At a finer time scale, small segments with different rate (slope)
can be identified (especially
during the first half, in the case of Fig.~\ref{fig:rates}(a)).
This feature is a manifestation of the change of rhythms throughout a match.
We estimated local rates $\lambda \equiv 1/\tau$ as $\lambda=n/t_{IN}$, where
$t_{IN}$ is the duration of each full sequence of ininterrupted touches
and $n$ is the number of touches in that sequence.
The histogram of rates $\lambda$ is shown in
Fig.~\ref{fig:rates}(b). Meanwhile  panel (c) displays  $\lambda$ as a
function of time, putting into evidence its fluctuating character.

\begin{figure}[t!]
\begin{center}
\includegraphics*[bb=100 480 560 750, width=0.45\textwidth]{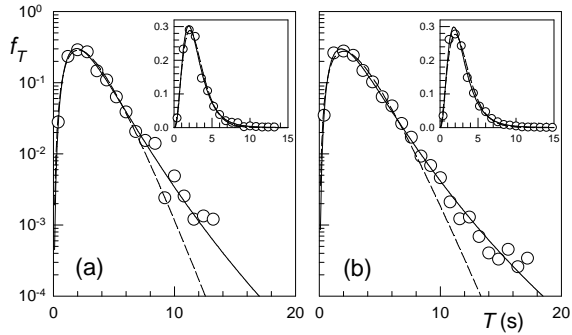}
\caption{
Normalized histograms of inter-touch times,
(a) for the match of Fig.~\ref{fig:series}  and  (b) for all the
matches investigated.
Dashed lines correspond to the gamma distribution with
parameters ($\beta,\tau/$s)= ($3.43\pm0.17$, $0.85\pm0.05$) (a) and
($3.12\pm 0.12$, $0.95\pm0.05$) (b).
Solid lines correspond to the $q$-gamma distribution with
parameters $(\beta,\tau/$s$,q$)= ($4.45\pm0.35$,$0.43\pm0.07,1.066\pm0.007$) (a),
($4.20\pm 0.17$, $0.43\pm0.04$,$1.076\pm0.003$) (b).
Insets: linear representations of the same data.
 }
\label{fig:qgamma}
\end{center}
\end{figure}

We will see that the fact that the rate of occurrence $\lambda$ is not constant,
but instead
it is a fluctuating quantity, may explain the behavior of the tails
of  inter-touch time distributions.
In effect, let us interpret the PDF $\Gamma_{\beta,1/\lambda}(t)$
as the conditional PDF $f_{T|\lambda}$ of variable $T$ given $\lambda$,
where $\lambda$ is a stochastic variable. Moreover, let us also assume
that $\lambda$ is gamma distributed, i.e.,
$f_{\lambda}(x)=\Gamma_{\alpha,\kappa}(x)$.
Although this may be only a crude estimation of the distribution of local
rates, it takes into account its main features,
except deviations at the tails (see Fig.~\ref{fig:rates}(b)).
Under the assumptions above,
the marginal PDF has the form \cite{beck1,beck0,beck2}
\begin{equation} \label{qgamma}
f_T(t)=
\int dx f_{\lambda}(x) f_{T|\lambda}(t,x)
={\cal N} t^{\beta-1}{\rm e}_q^{-t/{\tau}},
\end{equation}
where ${\cal N}$ is a normalization factor,
 $\tau=1/[(\alpha+\beta)\kappa]$,
$q=1+1/(\alpha+\beta)$ and the
$q$-exponential function (${\rm e}_q$) for negative argument is defined as
${\rm e}_q^{-x}=[1+(q-1)x]^\frac{1}{1-q}$,  if $q> 1$.
This PDF that generalizes the gamma distribution, is known as
$F$-distribution or also as
$q$-gamma distribution \cite{silvio}.
Panel (a) of Fig.~\ref{fig:qgamma} shows the WLS fit of the $q$-gamma function
to the same data of Fig.~\ref{fig:histos}.
In order to asses the goodness of fit we applied the Kolmogorov-Smirnoff test \cite{KS}.
We calculated confidence levels $\alpha$ by determining the largest deviation between the
cumulative distribution that arises from WLS fit and the observed one.
We obtained higher confidence levels for the $q$-gamma model. As illustration,
for match {\sf CB}, the gamma and $q$-gamma fits yielded $\alpha=7\%$ and 32\%,
respectively. Furthermore, the chi-square value of fit and the correlation coefficient
for match {\sf CB} were $(\chi^2, R^2$)= (0.0009, 0.99)
(against (0.002, 0.97) for the simple gamma distribution).

Although at the cost of introducing one more parameter,
the $q$-gamma model is satisfactory {\em for the full range} of values.
The advantage of introducing one more parameter was quantified through
Akaike information criterium (AIC=$2k-2\ln L$, where $k$ is the number of parameters and
$L$ the maximum likelihood) and also through Schwarz criterium (SIC=$k\ln(n)-2\ln L$,
where $n$ is the number of observations), which penalizes more strongly the introduction of
free parameters \cite{akaike}. In all cases the $q$-gamma distribution
yielded lower values.
For example,  (AIC, SIC)= (3726, 3741) (against (3760, 3772) obtained with
the simple gamma) for {\sf CB} and
(82694, 82718) (against (83650, 83666)) for the global set.

Furthermore the PDF of increments, which generalizes Eq. (\ref{teo}), namely,
\begin{equation}  \label{teo2}
f_{\Delta T}(t)\propto
\int_0^\infty [x(|t|+x)]^{\beta-1}{\rm e}_q^{-x/\tau} {\rm e}_q^{-(|t|+x)/\tau}\,dx\; \,,
\end{equation}
also describes better numerical data (see Fig.~\ref{fig:histos}(b), where
the full line is the predicted PDF of increments using the parameters of
Fig.~\ref{fig:qgamma}(a)).

In Fig.~\ref{fig:qgamma}(b), all the matches were merged.
In this case, the merging procedure by itself might give rise to the dispersion of
$\lambda$ responsible for the behavior of
the tails. This global analysis, however, is useful for characterizing the average
behavior of football activity as a whole.
Since, as observed before, diversity is not very high amongst matches,
one observes for the merged data a behavior similar to that
observed for a {\em single} match (illustrated in Fig.~\ref{fig:qgamma}(a)),
although the statistics is poorer in the later case.

The introduction of parameter $q$ allows to describe better the statistics of
rare events. Notice that, in comparison with the gamma fits,
the $q$-gamma ones yield $\beta$ about one unit larger and $\tau$ about one half smaller.
Alternatively, assuming  that the fluctuating rates obey
a gamma distribution $\Gamma_{\alpha,\kappa}$, one can obtain
$q$ (i.e., $q=1+1/(\alpha+\beta)$) and $\tau$ (i.e., $\tau=1/[(\alpha+\beta)\kappa]$) for
the resulting $q$-gamma distribution.
The values of $q$ are very close
to  those obtained by directly fitting the $q$-gamma distributions
($q\simeq 1.07$).
Whereas, the resulting values of $\tau$  are larger  than those obtained from
$q$-gamma fits but still of the order of 1s.
Therefore our model is selfconsistent.
The lack of a complete matching
of parameter $\tau$  is due to diverse reasons.
On one hand, there is a certain degree of arbitrariness in the definition of
local rates, that in our case were computed over each
continuous sequence of touches, through $n/t_{IN}$.
Moreover, the distribution of local rates as here defined are not strictly gamma,
but  only approximately.
Finally, also parameter $\beta$ is an averaged quantity
since the number of phases may fluctuate throughout a match.
Nevertheless, the comparison between theoretical and empirical distributions
supports the present model as a better approximation than the simple gamma distribution.

\section{In and out-of-play intervals}

\begin{figure}[ht]
\begin{center}
\includegraphics*[bb=60 440 550 720, width=0.46\textwidth]{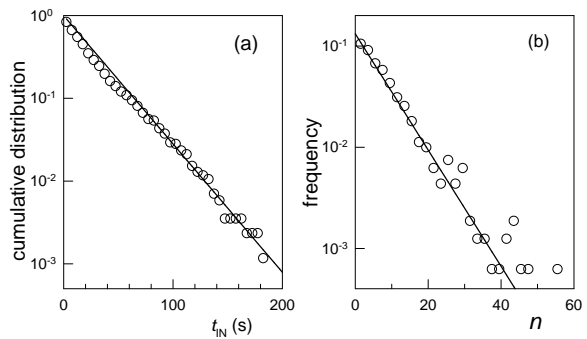}
\caption{ Statistics of in-play intervals.
Cumulative histogram of $t_{IN}$ (a).
The straight line corresponds to a exponential with characteristic time $\tau_{IN}=28 s$.
Normalized histogram of the number of touches in each continuous sequence (b).
}
\label{fig:tin}
\end{center}
\end{figure}

In Fig.~\ref{fig:tin}(a) we present the histogram of $t_{IN}$ (duration of the
intervals without interruption). Notice that $t_{IN}=\sum_{i=1}^{n}T_i$.
Then, its PDF can be obtained as
$\rho(t_{IN})=\sum_{n\ge 1}\rho(t_{IN}|n) P(n)$, where $P(n)$ is the distribution
of the number of touches $n$ in each continuous sequence. The
conditional PDF is in first approximation $\Gamma_{\beta,n\tau}$, assuming that
the $n$ inter-touch times are independent identically $\Gamma_{\beta,\tau}$
stochastic variables. On the other hand,
$P(n)$ follows approximately the exponential law ${\rm e}^{-n/n_o}/n_o$ (Fig.~\ref{fig:tin}(b)),
being $n_o \simeq 7.54\pm 0.21$ (from WLS fit), while $\langle n \rangle\simeq 8.4\pm 0.3$.
Under the assumptions above, one obtains a PDF that can be well approximated by an exponential
with characteristic time $\tau_{IN}\simeq \langle n \rangle \beta\tau$, consistent with the
numerical results in Fig.~\ref{fig:tin}(a).

\begin{figure}[h!b]
\begin{center}
\includegraphics*[bb=60 440 550 720, width=0.46\textwidth]{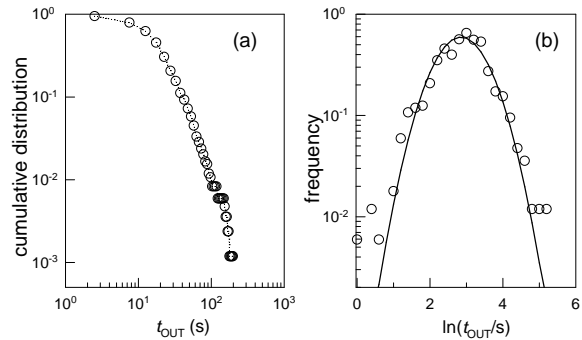}
\caption{ Statistics of out-of-play intervals.
Cumulative histogram of $t_{OUT}$ (a).
In (b), normalized histogram of $\ln t_{OUT}$ (symbols) and Gaussian fit with
mean 2.85 and variance 0.45 (solid line).}
\label{fig:tout}
\end{center}
\end{figure}

The histogram of times elapsed between sequences (duration of intervals
when the ball is out-of-play) was also computed and it is exhibited
in Fig.~\ref{fig:tout}(a).
Since both $t_{IN}$ and $t_{OUT}$ statistics are poor over a single match (about
one hundred events), we merged the records of several matches
({\sf AV,RA,MB,VZ,LF,BM,CB}, following Table~\ref{tableone}).
In these cases goal-keeper movements were computed.

Although the statistics of out-of-play intervals appears to display a power-law
behavior, a more careful analysis points to a log-normal statistics
(Fig.~\ref{fig:tout}(b)).
Up to now, the results were well understood within the framework of
non-homogeneous Poisson processes.
However, in the case of out-of-play intervals, the PDF basically
obeys a log-normal statistics that is not
straightforwardly related to the previous scenario.


\section{Final observations}

The statistics of touches can be understood on the basis
of Poissonian arrival or point processes.
Basically, events are not simply Poissonian but can be thought as composed
of different phases, as also observed for compound tasks. This explains the
behavior of the histogram of inter-touch times in the vicinity of the origin.
Meanwhile, non-homogeneities in the rate of occurrence (changes of rhythm)
throughout a match appear to be responsible for the power law tail in the distribution
of inter-touch times, giving rise to a $q$-gamma function.
It is noteworthy that a similar mechanism based on compound distributions
for the obtention of such PDFs
has been proposed within the context of the
``superstatistics''\cite{beck1}, where the fluctuating parameter is the temperature.
Here, we provide an example where $q$-gamma distributions
arise as a consequence of the fluctuating nature of a relevant parameter.

All the main features here exhibited and discussed for the matches
in Table \ref{tableone} are also
observed, in a preliminary analysis, for the sixty four matches of the 2006 World Cup
(results not shown).
Within the simple Poissonian framework,
the effective characteristic time  $\tau$ (of the order of one second) does not change
significantly from one match to another.
Meanwhile, parameter $\beta$ exhibits a greater variation, remaining approximately between
2 and 4, with average number of tasks close to 3.
In general there is a tendency that $\tau$ is shorter and $\beta$ larger
in more decisive matches or matches played by highly ranked teams.
These trends are qualitatively expected as far as in such matches players usually
save no efforts, playing faster and using more elaborated strategies.
In fact, larger $\beta$  already suggests more developed or complex actions.
The introduction of a further parameter, $q$,  which reflects the degree of
inhomogeneity of rhythms, improves the description of the tails.
The statistics of the duration of sequences of touches, interrupted by fouls,
ball leaving the field, etc., can also be derived within the same approach
used for the statistics of touches.

On the other hand, the statistics of intervals between sequences of touches
is of a different nature, belonging to the log-normal class.
There are diverse mechanisms that may give rise to such statistics. As an example,
for time series observed in turbulent flows, it has been attributed to a multiplicative
random process \cite{beck2}. However this issue
should be further investigated and deserves separate work.
\begin{widetext}

    \begin{table}
\begin{footnotesize}
\begin{tabular}{|cl|c||cl|c|}
  \hline
    &Match(cup) & Date &   &Match(cup) & Date \\
  \hline
  {\sf BR} &Barcelona 1$\times$1 R.Madrid(S)& 02/Apr./06  & {\sf PA} &S\~ao Paulo 4$\times$0 Atl\'etico-PR(L)& 14/Jul./05 \\
  {\sf RB} &R.Sociedad 0$\times$2 Barcelona(S)& 19/Mar./06& {\sf DC} &D. Cali 0$\times$1 Corinthians(L)& 15/Feb./06 \\
  {\sf AV} &A.Bilbao 1$\times$1 Villareal(S)& 26/Feb./06  & {\sf CU} &Corinthians 2$\times$2 U. Cat\'olica(L)& 22/Feb./06 \\
  {\sf RA} &R.Madrid 3$\times$0 Alaves(S)& 19/Feb./06      &{\sf PD} &Palmeiras 2$\times$0 D. T\'achira(L)& 25/Jan./06\\
  {\sf VB} &Valencia 1$\times$0 Barcelona(S)& 12/Feb./06  & {\sf PB} &Paran\'a 2$\times$0 Botafogo(B)& 03/Aug./05 \\
  {\sf MB} &Mallorca 0$\times$3 Barcelona(S)& 29/Jan./06   &{\sf PC} &Paran\'a 1$\times$2 Corinthians(B)& 14/Mar./06\\
  {\sf VZ} &Valencia 2$\times$2 Zaragoza(S)& 19/Jan./06   & {\sf CS} &Corinthians 1$\times$3 S\~ao Paulo(B)& 07/May/06 \\
  {\sf RM} &Reggina 1$\times$4 Milan (I)& 12/Feb./06      & {\sf SA} &Santos 2$\times$0 Atl\'etico-PR(B)& 23/Apr./06\\
  {\sf LF} &Leverkusen 2$\times$1 E. Frankfurt(G)& 28/Jan./06&{\sf SF} &S\~ao Paulo 1$\times$0 Flamengo(B)& 16/Apr./06 \\
  {\sf BM} &Borussia M. 1$\times$3 B. Munchen(G)& 27/Jan./06&{\sf PS} &P. Santista 0$\times$5 S\~ao Paulo(P)& 12/Feb./06\\
  {\sf CB} &Chelsea 2$\times$1 Barcelona (E)& 22/Feb./06   &{\sf SP} &Santos 1$\times$0 Palmeiras (P)& 05/Mar./06 \\
  {\sf AR} &Arsenal 0$\times$0 R. Madrid (E)& 08/Mar./06   &{\sf SC} &S. Caetano 2$\times$1 Corinthians(P)& 08/Feb./06\\
  {\sf SL} &S\~ao Paulo 1$\times$0 Liverpool(C)& 18/Dec./05&{\sf SM} &Santos 3$\times$2 Mar\'{\i}lia(P)& 22/Jan./06 \\
  \hline
\end{tabular}
\end{footnotesize}
\caption{Matches of South-American and European championships.
Cups are B (Brazilian), C (FIFA World Club), E (UEFA Champions
League), G (German), I (Italian), L (Libertadores),  P (Brazilian,
S\~ao Paulo State), S (Spanish). } \label{tableone}
\end{table}

\end{widetext}

\noindent
{\bf Acknowledgements: }
We thank Brazilian agency CNPq for partial financial support.
We also thank S. Picoli Jr. for interesting remarks.

\end{document}